\documentclass[
    ,final            
  ]
  {aipproc}

\layoutstyle{6x9}

\def\vect#1{{\mbox{\boldmath $#1$}}}

\def\Ahbar{ h \hspace{-2mm}\raisebox{1.3mm}{--}}

\begin{document}

\title{Systematic study of low-lying $E1$ strength using the time-dependent mean field theory}

\classification{21.60.Jz, 24.30.Gd}
\keywords{Hartree-Fock-Bogoliubov; linear response; deformed nuclei.}

\author{S. Ebata}{address={Center for Nuclear Study, Univ. of Tokyo, Bunkyo-ku, 113-0033, Japan}
,altaddress={Theoretical Nuclear Physics Laboratory, RIKEN Nishina Center, Wako-shi, 351-0198, Japan}
}

\author{T. Nakatsukasa}{address={Theoretical Nuclear Physics Laboratory, RIKEN Nishina Center, Wako-shi, 351-0198, Japan}
,altaddress={Center for Computational Sciences, Univ. of Tsukuba, Tsukuba-shi, 305-8571, Japan}
}

\author{T. Inakura}{ address={Theoretical Nuclear Physics Laboratory, RIKEN Nishina Center, Wako-shi, 351-0198, Japan}
}

\begin{abstract}
We carry out systematic investigation of electric dipole ($E1$) mode from light to heavy nuclei, 
using a new time-dependent mean field theory: the Canonical-basis 
Time-Dependent Hartree-Fock-Bogoliubov (Cb-TDHFB) theory.
The Cb-TDHFB in the three-dimensional coordinate space representation can deal with pairing correlation 
and any kind of deformation in the time-dependent framework. 
We report the neutron-number dependence of the low-energy $E1$ mode for light ($A<40$) and 
heavy isotopes ($A>100$) around $N=82$.
\end{abstract}

\maketitle


\section{Introduction}
Properties of the low-energy electric dipole ($E1$) excited states of neutron-rich nuclei are important to 
understand nucleosynthesis on the $r$-process path \cite{G1998}. 
These states are often called Pygmy dipole resonance (PDR). 
Due to the recent progress of radioactive facilities, we become able to generate and measure the neutron-rich 
nuclei, however it is still difficult to study nuclei on the $r$-process path. 
Thus, we strongly demand a method (model) 
that is able to analyze and predict the excitation modes of exotic nuclei 
in very neutron-rich and heavy nuclear region, such as nuclei on the $r$-process path. 
In order to study dynamical properties of heavy unstable nuclei, 
the method should take into account the effects of deformation
and pairing correlation. The Time-Dependent Hartree-Fock-Bogoliubov (TDHFB) or HFB plus
quasi-particle random-phase approximation (HFB+QRPA) are such candidates \cite{BR1986}, 
which are applicable to a wide range of nuclei from light- to heavy-mass regions. 
However, they require a significant effort for coding the program as well as large computational resources. 
The TDHFB calculations with full three-dimensional (3D) dynamics is currently still in a preliminary stage
\cite{HN2007,SB2011}. We propose a feasible approach, gCanonical-basis TDHFBh (Cb-TDHFB) 
in the 3D coordinate-space representation \cite{EN2010}. 
The numerical cost of Cb-TDHFB is much smaller than that of TDHFB, 
then we can study excited states of various isotopes systematically.

In this present work, 
we investigate the low-lying $E1$ modes of even-even neutron-rich nuclei using the Cb-TDHFB in the linear regime. 
We report that the low-lying $E1$ strengths of light and heavy neutron-rich isotopes 
have characteristic neutron-number dependence. 
\section{Formulation}
The Cb-TDHFB equations \cite{EN2010} are derived from the TDHFB equations with a simple approximation for the pairing interaction
which is using only diagonal part of pairing functional $\Delta_{lk} = \Delta_{l}\delta_{\bar{l}k}$ analogous to the BCS approximation. 
In present work, we choose a very schematic pairing functional as $E_{\rm Pair}=-\sum_{l,m>0}G_{lm} \kappa^{\ast}_{l}\kappa_{m}$ 
with $\kappa_{l}\equiv u_{l}v_{l}$. $u_{l}$ and $v_{l}$ correspond to the time-dependent BCS factors for the canonical pair of states, 
$\phi_l(\vect{r},\sigma,t)$ and $\phi_{\bar l}(\vect{r},\sigma,t)$.
The Cb-TDHFB equations for the canonical single-particle states, the occupation probabilities $\rho_{l}(t)\equiv |v_{l}(t)|^{2}$, 
and the pair probabilities $\kappa_{l}(t)$ are written as \\[-4mm]
\begin{eqnarray}
\left\{
\begin{array}{l}
 i\Ahbar \dot{\phi}_{l}(\vect{r},\sigma,t)=\big( \hat{h}[\rho(t)] - \varepsilon_{l}(t) \big) \phi_{l}(\vect{r},\sigma,t) , \quad
 i\Ahbar \dot{\phi}_{\bar l}(\vect{r},\sigma,t)=\big( \hat{h}[\rho(t)] - \varepsilon_{\bar l}(t) \big) \phi_{\bar l}(\vect{r},\sigma,t) , \\[1.5mm]
 i\Ahbar \dot{\rho}_{l}(t) = \kappa_{l}(t) \Delta^{\ast}_{l}(t) - \Delta_{l}(t) \kappa^{\ast}_{l}(t) , \\[1mm]
 i\Ahbar \dot{\kappa}_{l}(t)=(\varepsilon_{l}(t)+\varepsilon_{\bar{l}}(t))\kappa_{l}(t) +\Delta_{l}(t) (2\rho_{l}(t) -1) ,
\end{array}
\right. \nonumber
\end{eqnarray} \\[-2mm]
where, $\hat h[\rho(t)]$ is the HF Hamiltonian.
$\Delta_{l}(t)$ and $\varepsilon_{l}(t)$ are defined as $\Delta_{l}(t)\! \equiv\! \sum_{m>0}\! G_{lm} \kappa_{m}(t)$ and
$\varepsilon_{l}(t)\!\equiv\! \langle \phi_l(t) | \hat{h}(t) | \phi_l(t) \rangle$, respectively.
The Cb-TDHFB equations conserve the orthonormal property of canonical basis, 
the expectation value of particle number and total energy. They are identical to TDHFB.
We use the Skyrme functional of the SkM$^{\ast}$ parameter set for $ph$-channel  
and $G_{lm}=Gf(\varepsilon^{0}_{l})f(\varepsilon^{0}_{m})$, here $f(\varepsilon)$ is a cutoff function, 
adopted from Ref.\cite{TT1996} for $pp$,$hh$-channel. 

In the liner-response calculation, we add an external field which is weak and instantaneous in time,
$\hat{V}_{\rm ext}(\vect{r},t)=-k\hat{F}(r)\delta(t)$. $\hat{F}$ is one-body operator.
In this work, we adopt the $E1$ operator, 
$\hat{F} \equiv (Ne/A) \hat{\vect{r}}^{\rm (p)} - (Ze/A) \hat{\vect{r}}^{\rm (n)}$, 
where $r=(x,y,z)$ 
and ${\rm (p), (n)}$ mean the operation on protons and neutrons, respectively.
If the parameter $k$ is very small, the fluctuation of the nuclear density is linearized automatically, 
and this linear-response calculation is equivalent to the QRPA (or RPA). 
We obtain the strength function $S(\hat{F}; E)$ through the Fourier transformation of 
the expectation value of the one-body operator $\hat{F}$. 
Details can be found in Refs.\cite{EN2010,NY2005}.\\[-7mm]
\section{Results}
We discuss a systematic trend of the low-lying $E1$ strength. 
To quantify PDR, the following ratio of the low-lying $E1$ strength is used:
\begin{eqnarray}
\label{Ratio}
 \frac{m_{1}(E_{\rm c})}{m_{1}} \equiv \frac{\int^{E_{\rm c}} E\times S(E1; E) dE}{\int E\times S(E1; E) dE} \times 100 \ [\%],
\end{eqnarray}
where $E_{\rm c}$ is a cut-off energy. We take $E_{\rm c}=$10 MeV in the present calculation. 

Figure 1 shows the $\frac{m_{1}(E_{\rm c})}{m_{1}}$ for O, Ne and Mg isotopes 
as a function of neutron number. 
We see that the nuclei with $N=$8 - 14 have the ratio of low-lying $E1$ strength less than $1.0\%$, 
and a sudden jump of the ratio at $N=14\rightarrow 16$ on each isotopic chain. 
The neutron number $N=16$ corresponds to the occupation of the $s_{1/2}$ orbit.
These results of light neutron-rich nuclei suggest that the position of the neutron Fermi level plays an 
important role for the emergence of low-lying $E1$ strength \cite{IN2011}. 

Figure 2 shows the results for heavy neutron-rich isotopes which are $Z=38-54$ with $N=76-90$. 
Again we can see that a sudden jump of the ratio at $N=82 \rightarrow 84$ on each isotopic chain. 
These results show a neutron shell effect on the low-lying $E1$ strength. 
However, there are constant but significant PDR ratios for $N=76 - 82$. 
Although $^{130-136}$Xe are stable nuclei, they have $1.5\%$ of low-lying $E1$ strength. 
This ratio systematically increases as the proton number decreases. 
These features suggest that the PDR in heavy nuclei has a different structure from one of light nuclei. 
Currently, we are investigating the origin of the difference, 
and also studying even heavier neutron-rich nuclei around $N=126$.\\[-5mm]
\begin{center}
\begin{figure}[h]
   \includegraphics[keepaspectratio,width=50mm,angle=-90,clip]{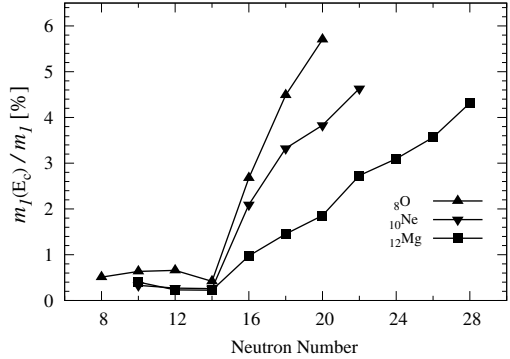}
\caption{
Neutron-number dependence of $m_{1}(E_{\rm c})/m_{1}$ defined in Eq.\eqref{Ratio} 
for O, Ne and Mg isotopes. }
\end{figure}\ \\[-0.5cm]
\begin{figure}[h]
   \includegraphics[keepaspectratio,width=50mm,angle=-90]{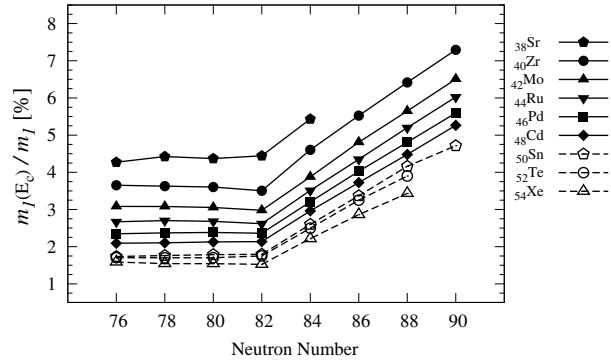}
\caption{
Same as Fig. 1 but for $Z=38-54$ (Sr to Xe) with $N=76-90$.}
\end{figure}
\end{center}
This work is supported in part by MEXT HPCI Strategic Program and by KAKENHI (Nos. 20165003 and 21340073). 

\bibliographystyle{aipproc}   

\end{document}